\documentclass[11pt,a4paper]{article}

\usepackage{epsfig,amsmath,amssymb}

\def\be{\begin{equation}}
\def\ee{\end{equation}}
\def\lp{\left(}
\def\rp{\right)}
\def\lb{\left[}
\def\rb{\right]}
\def\La{\Lambda}
\def\b2{\beta^2}

\begin{document}

\title{Thin-shell wormholes in Einstein--Maxwell theory with a Gauss--Bonnet term} 
\author{Marc Thibeault$^{1,}$\thanks{e-mail: marc@df.uba.ar}, Claudio Simeone$^{1,}$\thanks{e-mail: csimeone@df.uba.ar}\;  and Ernesto F. Eiroa$^{2,}$\thanks{e-mail: eiroa@iafe.uba.ar} \\
{\small $^1$ Departamento de F\'{\i}sica, Facultad de Ciencias Exactas y 
Naturales,} \\ 
{\small Universidad de Buenos Aires, Ciudad Universitaria Pab. I, 1428, 
Buenos Aires, Argentina} \\
{\small $^2$ Instituto de Astronom\'{\i}a y F\'{\i}sica del Espacio, C.C. 67, 
Suc. 28, 1428, Buenos Aires, Argentina}}

\maketitle

\begin{abstract}
We study five dimensional thin-shell wormholes in Einstein--Maxwell theory with a Gauss--Bonnet term. The linearized stability under radial perturbations and the amount of exotic matter are analyzed as a function of the parameters of the model. We find that the inclusion of the quadratic correction substantially widens the range of possible stable configurations, and besides it allows for a reduction of the exotic matter required to construct the wormholes. \\

\noindent 
PACS number(s): 04.20.Gz, 04.50.+h, 11.27.+d 

\end{abstract}

\section{Introduction}

Traversable Lorentzian wormholes are solutions of the equations of gravitation which connect two regions of the same universe, or of two universes, by a throat \cite{motho,visser}. The throat is defined as a minimal area hypersurface which satisfies a flare-out condition \cite{hovis1}. This requires the presence of exotic matter, that is, matter which violates the null energy condition (NEC) \cite{motho,visser,hovis1,hovis2}. Because it has been shown that the amount of exotic matter necessary for the existence of a wormhole can be made infinitesimally small by a suitable choice of the geometry \cite{viskardad}, considerable efforts have been addressed to precisely quantify such amount for different configurations, and to show how it can be minimized \cite{bavis}. Indeed, the total amount of exotic matter has been pointed as an indicator of the physical viability of traversable wormholes \cite{nandi2}.

Another central aspect of a wormhole --in fact, of any physically meaningful solution within any theory of gravitation-- is its stability. Within the framework of traversable wormholes, stability under perturbations preserving the symmetry of the original configuration has been widely analyzed. In particular, this problem has received considerable attention in the case of thin-shell wormholes, that is, wormholes which are mathematically constructed by cutting and pasting two manifolds to obtain a geodesically complete new manifold \cite{mvis}. In this case the exotic matter is located in a shell placed at the joining surface; the framework for dealing with these wormholes is the Darmois--Israel formalism, which leads to the Lanczos equations, that is, to the equations of gravitation projected on the joining surface \cite{daris,mus}. The solution of these equations, once provided an equation of state for the matter on the shell, determines the dynamical evolution. Such a procedure has been applied to spherically and cylindrically symmetric configurations associated to wormhole solutions within general relativity \cite{poisson,eilobo,barcelo}.  

The theory of gravity in five dimensions corresponding to the Einstein--Hilbert action supplemented with a Gauss--Bonnet term is, in a certain sense, the most general (metric) theory of gravity one can construct satisfying the conservation of the equations of motion which still remain being of second order \cite{Lovelock}. This theory, and its analogue in $D$ dimensions, was extensively studied in the last three decades and, in particular, the attention focused on it was mainly due to the fact that the theory arises within the string theoretical framework \cite{strings}. For instance, a version of this appears as corrections (proportional to the inverse of the string  tension) to the low energy effective action of the heterotic string theory \cite{G} as well as in Calabi--Yau compactification of the $M$-theory \cite{S}. Besides, these theories result closely related to  Chern--Simons gravity in odd dimensions which turns out to correspond to a particular choice of the parameters of the model; see Ref. \cite{BTZ}. Physically, adding the Gauss--Bonnet (higher order) terms in the gravitational action corresponds to the inclusion of short distance corrections to general relativity. The study of black hole solutions in Einstein--Gauss--Bonnet theory was initiated in the decade of 1980, when the statical spherically symmetric solution was reported by Boulware and Deser in Ref. \cite{BD}. Subsequently, Wiltshire derived in Ref. \cite{W} the charged black hole geometry in both Maxwell and Born--Infeld electrodynamics. Both geometrical and thermodynamical aspects of black holes are substantially modified by the addition of the Gauss--Bonnet term \cite{rafa}.

Lorentzian wormholes in spacetimes with more than four dimensions were analyzed by several researchers \cite{gbwh,hdwh}. In particular, wormholes in Einstein--Gauss--Bonnet gravity were considered in Ref. \cite {gbwh}. In the present work, the idea is then to use Lorentzian wormholes as a test bed to explore some of the qualitative changes that could happen in General Relativity with the addition of a Gauss-Bonnet term. More precisely, we shall study how the stability under radial perturbations of five dimensional spherically symmetric thin-shell wormholes in Einstein--Maxwell theory and the amount of exotic matter needed are affected by the presence of a Gauss-Bonnet term. In Section 2 we shall construct a generic thin-shell wormhole and write down the corresponding Lanczos equations; in Section 3 we shall analyze its mechanical stability under perturbations preserving the symmetry, and in Section 4  the energy conditions will be studied, and the total amount of exotic matter will be calculated. The dependence of the results in terms of the parameters of the model will be analyzed in detail. It will be shown that the inclusion of the Gauss--Bonnet term permits stability configurations with more physical values of $\beta^2$ with  small charge; also, we will see that the amount of exotic matter can be reduced for given values of the parameters. Section 5 is devoted to a brief summary and discussion.  Throughout the paper we set units so that $c=G=1$.

\section{Charged thin-shell wormholes}

The five-dimensional Einstein--Maxwell theory with a Gauss--Bonnet term representing a quadratic curvature correction is given by \cite{BD,W}
\be
S=\int d^{5}x\sqrt{-g}\lb R-2\La-\frac{1}{4}F_{\mu\nu}F^{\mu\nu}+\alpha\lp R_{\alpha\beta\gamma\delta}R^{\alpha\beta\gamma\delta}-4R_{\alpha\beta}R^{\alpha\beta}+R^2\rp\rb,
\ee
where  the signature chosen is $(-++++)$, $\La$ is the cosmological constant, and $\alpha$ is a constant of dimensions $(\mathrm{length})^2$. The variational principle $\delta S=0$ leads to the Einstein--Maxwell equations
\be
R_{\mu\nu}-\frac{1}{2}g_{\mu\nu}R+\La g_{\mu\nu}=\frac{1}{2}\lp T_{\mu\nu}^{EM}+T_{\mu\nu}^{GB}\rp,
\ee
\be
T_{\mu\nu}^{EM}=  F_{\mu \alpha}F_{\nu }^{\; \alpha}-\frac{1}{4}g_{\mu \nu }F_{\alpha\beta}F^{\alpha\beta},
\ee
\be
T_{\mu\nu}^{GB}=\alpha\lb 8R_{\alpha\beta}R^{\alpha\ \beta}_{\ \mu\ \nu}-4R_{\mu\alpha\beta\gamma}R_\nu^{\ \alpha\beta\gamma}+8R_{\mu\alpha}R^\alpha_{\ \nu}-4RR_{\mu\nu}+g_{\mu\nu} \lp R_{\alpha\beta\gamma\delta}R^{\alpha\beta\gamma\delta}-4R_{\alpha\beta}R^{\alpha\beta}+R^2\rp\rb,
\ee
where $T_{\mu\nu}^{EM}$ is the usual electromagnetic energy-momentum tensor, and $T_{\mu\nu}^{GB}$ is an effective tensor associated with the quadratic Gauss--Bonnet term included in the action. These equations admit a spherically symmetric solution given by \cite{W}
\be 
ds^2=-f(r)dt^2+f^{-1}(r)dr^2+r^2(d\theta ^2+\sin^2\theta d\chi^2+\sin^2\theta\sin^2\chi d\varphi^2), 
\label{metric1}
\ee
\be
f(r)=1+\frac{r^2}{4\alpha}-\frac{r^2}{4\alpha}\sqrt{1+\frac{16M\alpha}{\pi r^4}-\frac{8Q^2\alpha}{3r^6}+\frac{4\La\alpha}{3}}.\label{metric2}
\ee
The non null components of the electromagnetic tensor in an orthonormal frame  are $F_{\hat{t}\hat{r}}=-F_{\hat{r}\hat{t}}=Q/4\pi r^{3}$. It is not difficult to check that  in the limit $\alpha\to 0$ the five dimensional Einstein--Maxwell solution with cosmological constant is recovered. In this limit,  for $\La=0$ the five dimensional Reissner--Nordstr\"{o}m  metric is obtained, so $M>0$  and $Q$ can be identified with the mass and charge respectively. For $\alpha \neq 0$, there is a minimum value of the radial coordinate $r_{min}$ such that the function inside the square root in Eq. (\ref{metric2}) is positive for $r> r_{min}$, so the metric (\ref{metric1}) is well defined. The geometry has a curvature singularity at the surface defined by $r=r_{min}$ \cite{W}. Depending on the values of the parameters, this singular surface can be surrounded by an event horizon with radius $r_{h}$, so the metric (\ref{metric1}) represents a black hole, or in absence of the event horizon, it is a naked singularity.

We shall construct a  spherically symmetric thin-shell wormhole starting from the generic geometry (\ref{metric1}), and introduce the explicit form (\ref{metric2}) in the final results. We take two copies of the region $r\geq a$, with $a$ greater than $r_h$ and $r_{min}$ to avoid possible horizons and singularities in our geometry, and paste them to obtain a geodesically complete new manifold with a matter shell at the surface $r=a$, where the throat of the wormhole is located. The procedure follows the steps of the Darmois-Israel formalism; in terms of the original coordinates $X^\gamma=(t,r,\theta,\chi,\varphi)$, on the shell we define  the coordinates $\xi^i=(\tau,\theta,\chi,\varphi)$, with $\tau$ the proper time. Thus, using an orthonormal basis $\{ e_{\hat\tau},e_{\hat\theta},e_{\hat\chi},e_{\hat\varphi}\}$, the extrinsic curvature at the two sides of the shell reads
 \be
K_{\hat i\hat j}^{\pm} = - n_{\gamma}^{\pm} \left. \left( \frac{\partial^2
  X^{\gamma}}{\partial \xi^{\hat i} \partial \xi^{\hat j}} + \Gamma_{\alpha \beta}^{\gamma}
  \frac{\partial X^{\alpha}}{\partial \xi^{\hat i}} \frac{\partial
  X^{\beta}}{\partial \xi^{\hat j}} \right) \right|_{r=a}, \label{e6}
\ee
where $ n_{\gamma}^{\pm}$ are the unit normals to the surface. Defining
$
\kappa_{\hat i\hat j}=K_{\hat i\hat j}^+-K_{\hat i\hat j}^-
$
 and $\kappa=tr(\kappa_{\hat i\hat j})$, we obtain the  Lanczos equations (the Einstein's equations on the shell)
\be
\kappa_{\hat i\hat j}-\kappa g_{\hat i\hat j}=-8\pi S_{\hat i\hat j},
\ee
where
$
g_{\hat i\hat j}=\mathrm{diag} (-1,1,1,1)
$ and $S_{\hat i\hat j}=\mathrm{diag}(\sigma,p_{\hat\theta},p_{\hat\chi},p_{\hat\varphi})$ is the surface energy-momentum tensor. To allow for the analysis of radial perturbations, we let the throat radius to vary with the proper time: $a=a(\tau)$. As a consequence of  the generalized Birkhoff theorem proved in Ref. \cite{W}, the geometry will remain given by (\ref{metric1}) and (\ref{metric2}) for any $r$ greater than $a(\tau)$.  
The resulting  expressions for the energy density and pressures for a generic metric function $f$ turn to be
\be
S_{\tau\tau}=\sigma=-\frac{3}{4\pi a}\sqrt{f(a)+\dot a^2}\label{sigma},
\ee
\be
S_{\hat\theta\hat\theta}=S_{\hat\chi\hat\chi}=S_{\hat\varphi\hat\varphi}=p=-\frac{2}{3}\sigma+\frac{1}{8\pi}\lp\frac{2\ddot a+f'(a)}{\sqrt{f(a)+\dot a^2}}\rp ,\label{p}
\ee
where the overdot and the prime means, respectively, the derivatives with respect to $\tau$ and $r$. As it was to be expected, the energy density is negative, revealing the existence of exotic matter at the shell. It is easy to see from Eqs. (\ref{sigma}) and (\ref{p})  that the following conservation equation is fulfilled:
\be
\frac{d}{d\tau}\lp\sigma a^3\rp+p\frac{d}{d\tau}\lp a^3\rp=0.\label{cons}
\ee
For a static configuration of radius $a_0$ we simply have
\be
\sigma_0=-\frac{3}{4\pi a_0}\sqrt{f(a_0)},\ \ \ \ \ \ \ \ \ p_0=-\frac{2}{3}\sigma_0+\frac{1}{8\pi}\lp\frac{f'(a_0)}{\sqrt{f(a_0)}}\rp.\label{eq}
\ee
Note that as the wormhole radius approaches  which would be the event horizon radius $r_h$ in the original metric, the energy density approaches to zero, but, instead, the pressure diverges unless $f'(r_h)=0$; this will be discussed in detail in section 4.

\section{Stability analysis}

We shall study the stability of the configuration under small perturbations preserving the symmetry; for this we shall follow the procedure first applied to thin-shell wormholes in Ref. \cite{poisson}. The dynamics of the shell results  from the Eqs. (\ref{sigma}) and (\ref{p}) or, alternatively, from one of them and the conservation equation (\ref{cons}). In any case an equation of state for the matter on the shell must be provided; because we are interested in studying small radial perturbations around a  radius of equilibrium $a_0$, we propose a linear relation 
\be
p=p_0+\b2(\sigma-\sigma_0),
\ee
where $\sigma_0$ and $p_0$ are the energy density and pressure corresponding to the equilibrium radius of the wormhole;
for ordinary matter $\beta^2$ would be the speed of sound, but due to the presence of exotic matter we shall regard it as an arbitrary constant.
 
These equations lead to an explicit relation between the energy density and the radius:
\be
\sigma (a)=\lp \frac{\sigma _{0}+p_{0}}{ \b2 +1}\rp \lp \frac{a_{0}}{a}\rp ^{3(\b2 +1)}+\frac{\b2 \sigma _{0}-p_{0}}{\b2 +1}.
\ee
Introducing this in Eq. (\ref{sigma}) we obtain the equation of motion 
\be
\dot a^2+V(a)=0,
\ee
where the potential $V(a)$ is defined as
\be
V(a)=f(a)-\frac{16\pi^2}{9} a^2\sigma^2.
\ee
It is easy to verify that the potential fulfils
$V(a_0)=V'(a_0)=0, $ so that the stable equilibrium configurations correspond to the condition $V''(a_0)>0$. The second derivative of the potential for the generic metric (\ref{metric1}) is given by
\begin{eqnarray}
V''(a_0)& = & f''(a_0)-\frac{32\pi^2}{9}\lb\sigma_0^2+4a_0\sigma_0\sigma'_0+a_0^2\lp{\sigma'_0}^2+\sigma_0\sigma''_0\rp\rb\nonumber\\ & = & f''(a_0)-\frac{32\pi^2}{9}\lb\sigma_0^2+9(\sigma_0+p_0)^2+9\b2\sigma_0(\sigma_0+p_0)\rb\nonumber\\
& = & f''(a_0)+\lp\frac{3\b2+2}{a_0}\rp\lb f'(a_0)-\frac{2f(a_0)}{a_0}\rb-\frac{f'^2(a_0)}{2f(a_0)}.
\end{eqnarray}
Hence the stability of the configuration requires the relation
\be
\b2>\frac{a_0^2f'^2(a_0)-2a_0^2f''(a_0)f(a_0)}{6a_0f'(a_0)f(a_0)-12f^2(a_0)}-\frac{2}{3}\label{18a}
\ee
if $f'(a_0)-2f(a_0)/a_0>0$, and
\be
\b2<\frac{a_0^2f'^2(a_0)-2a_0^2f''(a_0)f(a_0)}{6a_0f'(a_0)f(a_0)-12f^2(a_0)}-\frac{2}{3}\label{18b}
\ee
if $f'(a_0)-2f(a_0)/a_0<0$. Although it is possible to write explicitly an analytic expression of $\b2$ as a function of the parameters, the complexity of the formulas (\ref{18a}) and (\ref{18b}) inhibits to get a clear insight of the stability regions. It is thus preferable, as it is  customary, to draw the curves $V''(a_0)=0$, which allow for  an intuitive understanding of the behavior of  the configuration.
 We shall consider the cases $\La=0$ and $\La\neq 0$ separately, and thoroughly discuss the dependence of the  regions of stability  with different choices of the parameters.

\begin{figure}[t!]
\begin{center}
\vspace{0cm} 
\includegraphics[width=12cm]{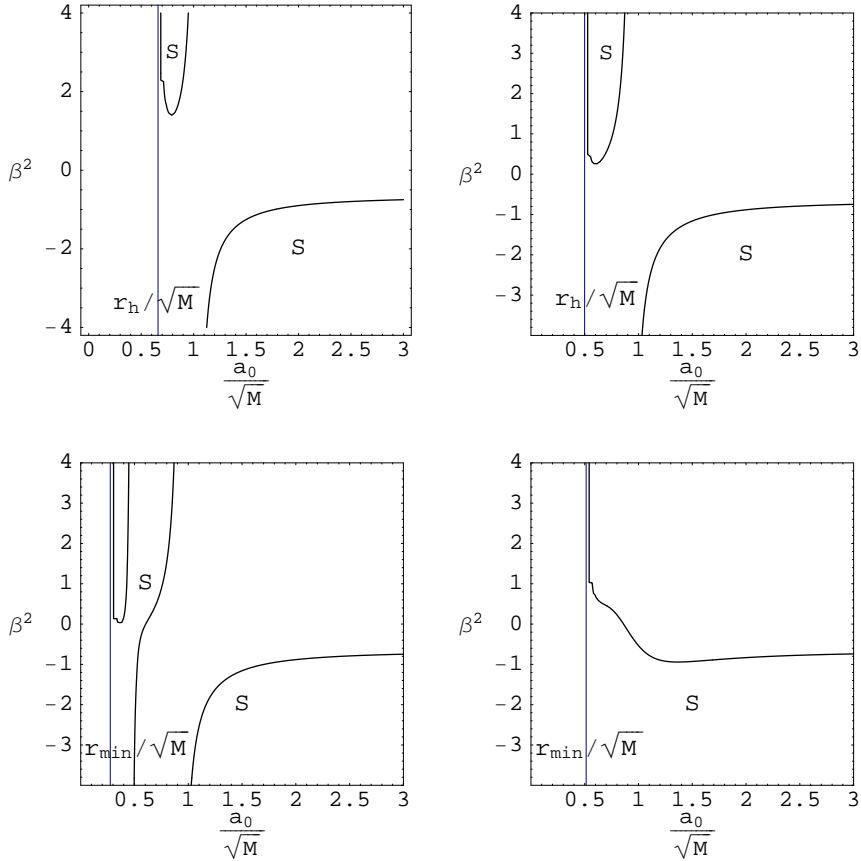} 
\vspace{-0.5cm}
\end{center}
\caption{The stability regions (marked with an S) are shown for  $\La=0$ and $\alpha/M=0.1$, which implies $|Q_c|/M=0.38.$ Only physically admissible regions, $r>r_{h}$ or $r>r_{min}$, are considered (see the text). For the upper left figure $|Q|=0$; in the upper right figure, $|Q|=0.99 |Q_c|$; in the lower left, $|Q|=1.01|Q_c|$; and for the last one $|Q| = 2|Q_c|$.}
\label{f1}
\end{figure}

\begin{figure}[t!]
\begin{center}
\vspace{0cm} 
\includegraphics[width=12cm]{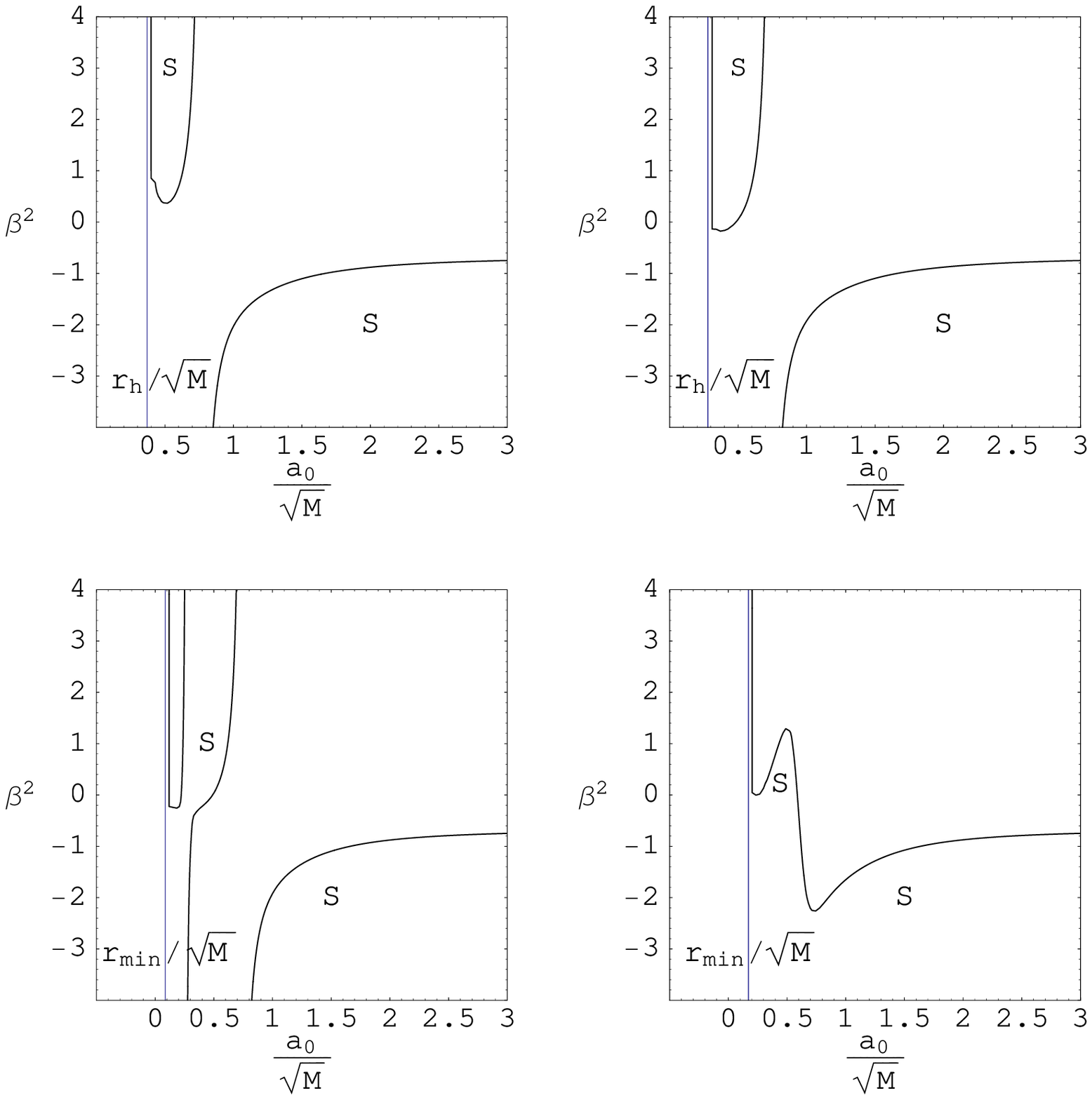} 
\vspace{-0.5cm}
\end{center}
\caption{The stability regions  are shown for  $\La=0$ and $\alpha/M=0.25$, which implies $|Q_c|/M=0.12.$ For the upper left figure $|Q|=0$; in the upper right figure, $|Q|=0.99 |Q_c|$; in the lower left, $|Q|=1.01|Q_c|$; and for the last one $|Q| = 2|Q_c|$.}
\label{f2}
\end{figure}

\begin{figure}[t!]
\begin{center}
\vspace{0cm} 
\includegraphics[width=12cm]{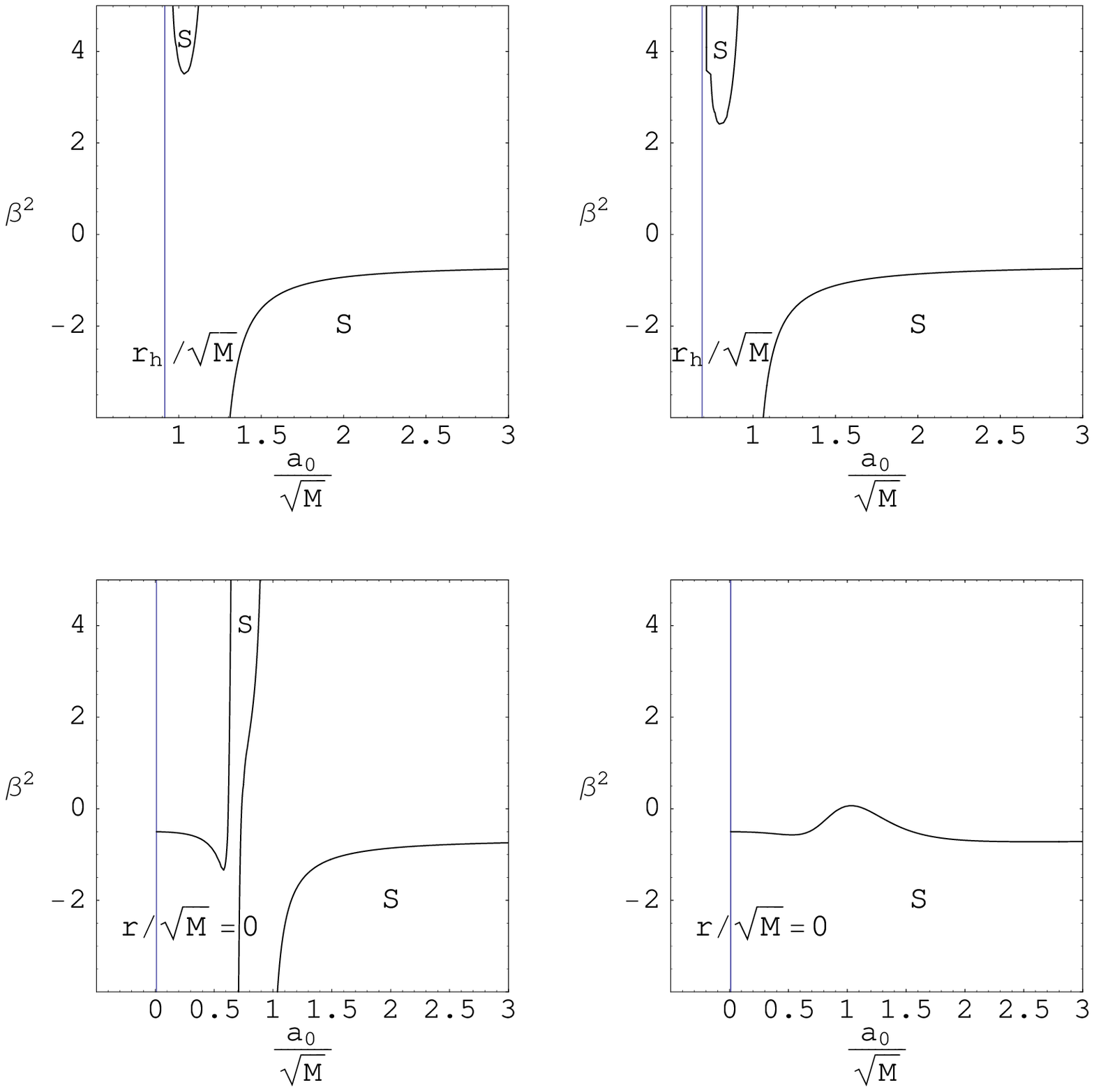} 
\vspace{-0.5cm}
\end{center}
\caption{The stability regions are shown for $\La=0$ and $\alpha/M=-0.1$, which implies $|Q_c|/M=0.725.$ For the upper left figure $|Q|=0$; in the upper right figure, $|Q|=0.99 |Q_c|$; in the lower left, $|Q|=1.01|Q_c|$; and for the last one $|Q| = 2|Q_c|$.}
\label{f3}
\end{figure}

\begin{figure}[t!]
\begin{center}
\vspace{0cm} 
\includegraphics[width=12cm]{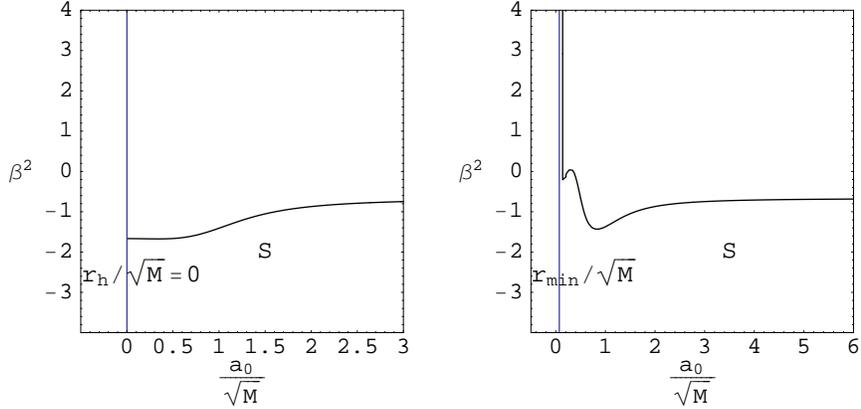} 
\vspace{-0.5cm}
\end{center}
\caption{The stability regions are shown for $\La=0$ and $\alpha/M=1/\pi$, which implies $|Q_c|/M=0$. The left figure corresponds to $|Q|/M=0$, and the right one to $|Q|/M=0.1$.}
\label{f4}
\end{figure}

\begin{figure}[t!]
\begin{center}
\vspace{0cm} 
\includegraphics[width=12cm]{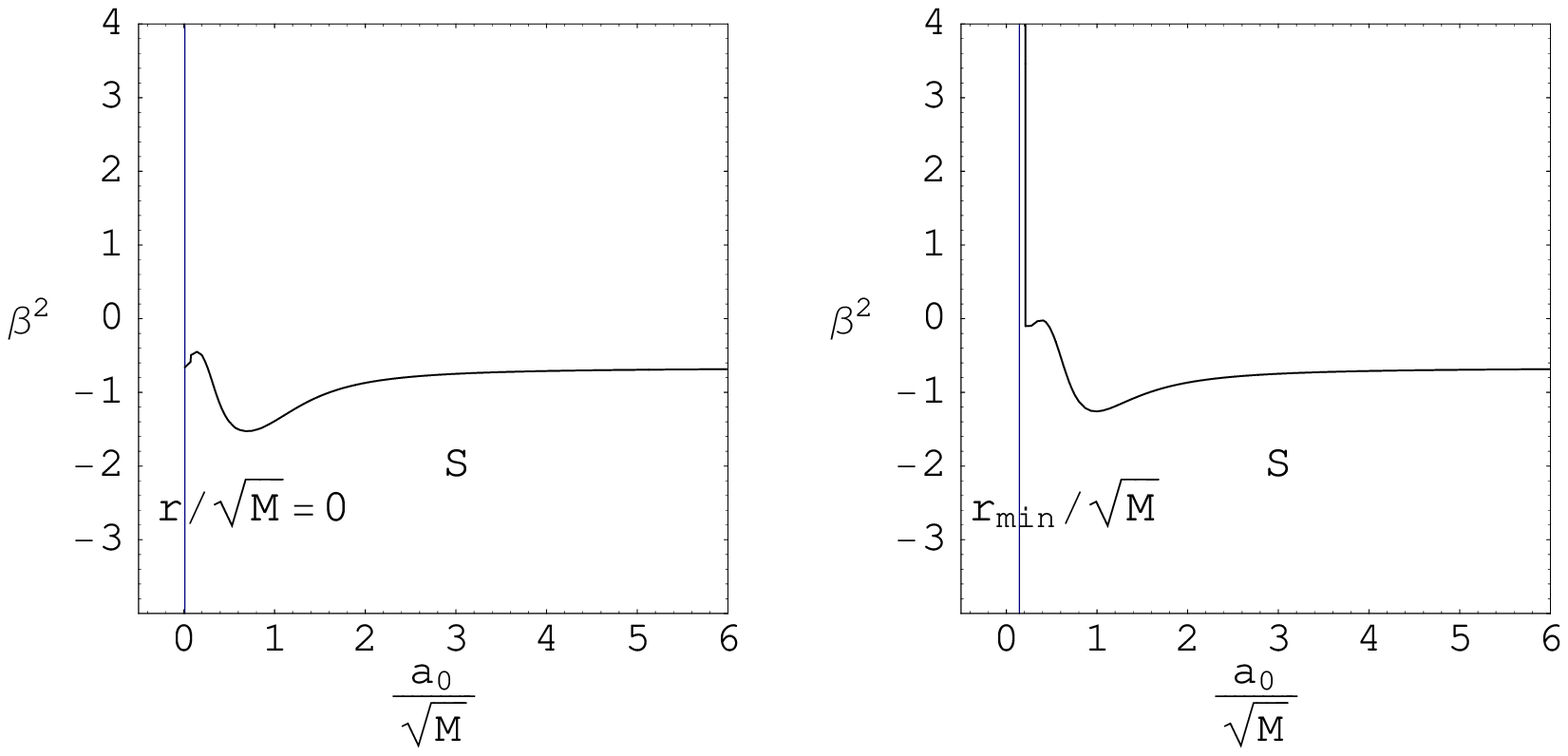} 
\vspace{-0.5cm}
\end{center}
\caption{The stability regions are shown for $\La=0$ and $\alpha/M=1.01/\pi.$ The left figure corresponds to $|Q|=0$ and the other one to $|Q|/M=0.2 $.}
\label{f5}
\end{figure}

\subsection{Case $\La =0$}

As pointed above, the square root of the  parameter  $\alpha$ of the theory introduces a length scale such that the Gauss--Bonnet corrections become relevant when the typical dimensions  of a given configuration are of order $\sqrt{|\alpha|}$. Thus, it is natural to perform an analysis for different  cases depending on the relation existing between $\alpha$ and the mass $M$. 

In the case of null cosmological constant the metric from which we start presents a singularity located at $r_{min}$ given by the greatest real and positive solution of the equation 
\be
r^6+\frac{16M\alpha}{\pi}r^2-\frac{8Q^2\alpha}{3}=0.\label{rmin0}
\ee
If Eq. (\ref{rmin0}) has no real positive solutions we have $r_{min}=0$, where the metric diverges. This singularity is surrounded, in principle, by two horizons with radii  
\be
r_{\pm}=\left\{\frac{M}{\pi}-\alpha\pm\lb\lp\frac{M}{\pi}-\alpha\rp^2-\frac{Q^2}{3}\rb^{1/2}\right\}^{1/2}.
\ee
The event horizon is placed at $r_h=r_+$, and $r_-$ is the inner horizon. For $\alpha>-M/\pi$, $r_{min}<r_h$ and the singularity can be shielded by the event horizon. But when $\alpha \leq -M/\pi $, we have a naked singularity because  $r_{min}\geq r_h$. For $|\alpha|<M/\pi$, it is easy to see that there exists a critical value of the charge 
\be
|Q_c|=\sqrt{3}\left|\frac{M}{\pi}-\alpha\right|,
\ee
such that if $|Q|<|Q_c|$ there are two horizons, if $|Q|=|Q_c|$ there is only one (degenerate) horizon, and if  $|Q|>|Q_c|$ there are no horizons. For $|\alpha|\geq M/\pi$, no horizons exist for any value of the charge (except the nonphysical solution $r_h=0$ for $\alpha=M/\pi$ and $Q=0$).  As mentioned in Section 2, the wormhole radius $a_0$ is taken greater than $r_h$ to avoid the presence of event horizons in our wormhole geometries. Note that when there is no horizon, the presence of the singular surface in $r=r_{min}$ compels us to consider values of $a_0$ greater than $r_{min}$  for the radius of  the wormhole throat.  

The associated stability analysis reveals both analogies and remarkable differences with the general relativity case corresponding to $\alpha\to 0$: 1) Two distinct regimes take place for   $|\alpha|<M/\pi$ and $|\alpha| \geq M/\pi$. In the first range most relevant results regarding the stability of the solutions appear, while in the second one stability requires $\beta^2<0$ (Figs. \ref{f4} and \ref{f5}). 2) For $\alpha >0$ the critical value of charge, $|Q_c|$, is smaller than the value corresponding to general relativity; instead for  $\alpha <0$ the critical charge is larger than in the absence of the Gauss--Bonnet terms. For  $\alpha \to 0$ (five dimensional  Reissner--Nordstr\"{o}m metric) the critical  value of the charge is  $|Q_c|=\sqrt{3} M/\pi$. 3) As  larger values of $\alpha$ are considered, the  regions of stability become enlarged (including $\beta^2<1$, which is an interesting feature) without the necessity of large values of the charge; in particular (see upper left in Fig. 2), this is possible with zero charge, which constitutes a drastic difference comparing with the general relativity case. Besides, for  $|Q|\gtrapprox |Q_c|$ we find a range of radii $a_0$ for which stability is achieved with any value of the parameter $\beta^2$ (lower left in Fig. 2).   4) For  $-M/\pi <\alpha <0$ the regions of stability turn to be smaller than without the Gauss--Bonnet quadratic contribution (see Fig. 3).

\subsection{Anti--De Sitter case}

 The presence of the cosmological constant introduces a restriction on the admissible values for $\alpha$: it must be $\La\alpha>-3/4$, in order to keep the metric real for large values of $r$. Now the singular surface radius $r_{min}$ is given by the greatest real and positive solution of the equation
\be
\lp 1+\frac{4\La \alpha }{3}\rp r^6+\frac{16M\alpha}{\pi}r^2-\frac{8Q^2\alpha}{3}=0, \label{rmin1}
\ee
or $r_{min}=0$ if all real solutions are non positive. The addition of the cosmological constant makes more complicated the structure of the horizons in the original manifold. Indeed we now have that the horizons are real and non-negative solutions of the equation
\be
\La r^6-6r^4+12\lp \frac{M}{\pi}-\alpha\rp r^2-2Q^2=0,\label{cubic}
\ee
subject to the constraints $r_h>r_{min}$ and $r_h^2\geq -4\alpha$. For $\alpha\leq 3\lb -1+\sqrt{1-16M\La /(3\pi)}\rb /(8\La)$, we have a naked singularity because $r_h\leq r_{min}$. When $3\lb -1+\sqrt{1-16M\La /(3\pi)}\rb /(8\La)<\alpha < M/\pi$, there is a critical value of charge now given by
\be
|Q_c|=\frac{2}{|\La |}\left\{-2+3\La \lp \frac{M}{\pi}-\alpha \rp +2\lb 1-\La \lp \frac{M}{\pi}-\alpha \rp \rb ^{3/2}\right\}^{1/2}.
\label{qcl}
\ee
The number of horizons are two when $|Q|<|Q_c|$, one (degenerate) for $|Q|=|Q_c|$ and zero if $|Q|>|Q_c|$.  When $\alpha >M/\pi$ we have again a naked singularity. As it was previously said the value of $a_0$ is taken greater than $r_{min}$ and $r_h$.

The critical value of charge for fixed $\alpha$ is an increasing function of $\La$, then for $\La<0$ its value is smaller than for $\La=0$.  For small values of $|\La|$, which are the most interesting ones from a physical point of view, a numerical calculation shows that the stability regions slightly change compared with the case $\La=0$. For the sake of brevity, the plots are not included.

\subsection{De Sitter case}

The case $\La> 0$ is considerably different from the point of view of the character of the horizons in the original manifold: for $\La> 0$ a cosmological horizon exists. Therefore the shell should always be placed inside the cosmological horizon. The positions of the singular surface and the horizons are obtained again from Eqs. (\ref{rmin1}) and (\ref{cubic}) respectively. When $0<\La \leq 3\pi/(16M)$ we have that $r_{min}\geq r_h$ for $3\lb -1-\sqrt{1-16M\La /(3\pi)}\rb /(8\La)\leq \alpha\leq 3\lb -1+\sqrt{1-16M\La /(3\pi)}\rb /(8\La)$, so there is a naked singularity. With other combinations of the parameters it is  $r_{min}< r_h$, if $r_h$ exists. For $M/\pi -1/\La <\alpha <M/\pi $, besides the cosmological horizon,  there are two, one or zero additional  horizons for $|Q|<|Q_c|$, $|Q|=|Q_c|$ and $|Q|>|Q_c|$, respectively, with the critical value of charge  again given by Eq. (\ref{qcl}). When $\alpha <M/\pi -1/\La $ or $\alpha >M/\pi $ there are no horizons and we have again a naked singularity. As in the other cases, the value of $a_0$ is taken greater than $r_{min}$ and $r_h$.

As in the preceding case, $|Q_c|$ increases with $\La$, so that the critical value in  De Sitter case is greater than for $\La=0$. Now there is a change in the values of $\alpha $ for which the largest regions of stability are found, namely $M/\pi -1/\La <\alpha <M/\pi $.  However, the stability analysis for small $\La$ does not reveal any remarkable aspect differing from the case $\La=0$, except  that the stability region corresponding to $\beta^2<0$ becomes limited for large  values of the wormhole radius due to the presence of the cosmological horizon. Again, for the same reason as in the Anti-De Sitter case, the plots are omitted.

\section{Energy conditions and exotic matter}

Quantifying the amount of exotic matter has been considered as a way to characterize  the viability of traversable wormholes \cite{nandi2}. Here we shall analyze the energy conditions and evaluate the total amount of exotic matter for the wormholes constructed in Section 2, in the case of static configurations, i.e. for $a=a_0$. 

The weak energy condition (WEC) states that for any timelike vector $u^\mu$ it must be $T_{\mu \nu}u^{\mu }u^{\nu }\ge 0$; the WEC also implies, by continuity, the null energy condition (NEC), i.e. that for any null vector $k^\mu$ it must be $T_{\mu \nu}k^{\mu }k^{\nu }\ge 0$ \cite{visser}. In an orthonormal basis the WEC reads $
\rho \ge 0,\ 
\rho +p_{j}\ge 0\  \forall j$, while the NEC takes the form $
\rho +p_{j}\ge 0\ \forall j$. In the case of the wormhole constructed above with radial pressure $p_r=0$, we have $\sigma <0,\ \sigma +p_{r}<0$, so that both energy conditions are violated. The transverse pressure is $p_t=p$ and the sign of $\sigma +p_{t}$, instead, is not fixed, but depends on the values of the parameters.

There have been several proposals for quantifying the  amount of exotic matter in the wormhole; two of them are the integrals \cite{viskardad,bavis} 
\be
\int \rho \sqrt{-g}\, d^4x,
\ \ \ \ \ \ \ \ \ 
\int (\rho+p_{i}) \sqrt{-g}\, d^4x,
\ee 
where $g$ is the determinant of the metric tensor. The most usual choice is the integral including the pressure associated to the violation of the energy conditions:
\be
\Omega =\int (\rho+p_{r}) \sqrt{-g}\, d^4x.
\ee
In our case, introducing the new coordinate $
{\cal R}=\pm(r-a_0)
$ with $\pm$ corresponding to each side of the shell, we have
\be
\Omega=\int\limits_{0}^{2\pi}\int\limits_{0}^{\pi}\int\limits_{0}^{\pi}
\int\limits_{-\infty}^{\infty}(\rho+p_{r}) \sqrt{-g}\, 
d{\cal R}d\theta\, d\chi\, d\varphi.
\ee
Because the shell does not exert radial pressure, and the energy density is located on a surface, so that $
\rho =\delta({\cal R})\sigma_0 
$, then we simply have
\be
\Omega = \int\limits_{0}^{2\pi}\int\limits_{0}^{\pi}\int\limits_{0}^{\pi} \sigma \left.
\sqrt{-g}\right|_{r=a_0}\,d\theta\,d\chi\, d\varphi=4\pi^2 a_0^3\sigma_0 .
\ee
Thus we find that
\be
\Omega = -3\pi a_0^2\sqrt{f(a_0)}.
\ee
Replacing the explicit form of the metric (\ref{metric2}), simple expressions for the behavior of $\Omega$ with the  wormhole radius can be obtained for the limiting case $a\to\infty$, which makes sense only for $\Lambda \leq 0$ and $\Lambda\alpha>-4/3$. For $\Lambda=0$ we have 
\be
\Omega \approx  -3\pi a_0^2.
\ee
For $\Lambda<0$ the limiting expressions differ depending on the sign of $\alpha$; for $\alpha>0$ we obtain
\be
\Omega \approx  -\frac{3\pi a_0^3}{2\sqrt{\alpha}}\lb 1-\sqrt{1+\frac{4\Lambda\alpha}{3}}\rb^{1/2},
\ee
while for $\alpha<0$ we have
\be
\Omega \approx  -\frac{3\pi a_0^3}{2\sqrt{|\alpha|}}\lb \sqrt{1+\frac{4\Lambda\alpha}{3}}-1\rb^{1/2}.
\ee

\begin{figure}[t!]
\begin{center}
\vspace{0cm} 
\includegraphics[width=16cm]{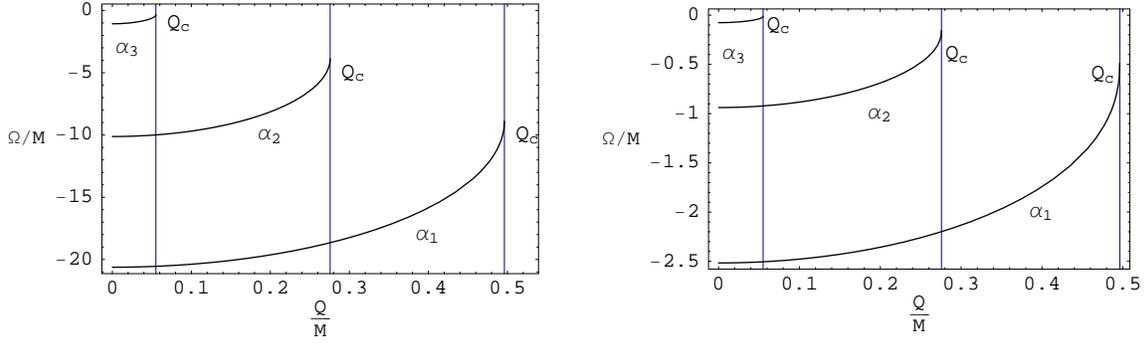}
\vspace{-0.5cm}
\end{center}
\caption{Amount of exotic matter (for $\La= 0$) as a function of the charge, for different values of the parameter $\alpha$ and the wormhole radius $a_0$. Left: $a_0=2.1r_h$; right: $a_0=1.1r_h$. In both cases,  $\alpha_1=0.1M/\pi$, $\alpha_2=0.5M/\pi$, and $\alpha_3=0.9M/\pi$. The horizon radius depends on the value of $\alpha$. Note the difference in the scales in the energy axis between the two graphics.}
\label{en}
\end{figure}

A natural question is which are the conditions such that the amount of exotic matter can be reduced. Because $\Omega$ is proportional to $\sigma_0$, which, as anticipated in Section 2, approaches to zero when the wormhole radius tends to the event horizon of the original metric, we shall analyze this limit in detail. We consider values of $\alpha$ for which horizons effectively exist; the number of horizons in the original manifold  is in this case determined by the value of the charge. When  $|Q|$ is less that the critical value of charge $|Q_c|$ defined in Section 3, there are two horizons  (within this analysis we are not interested in the cosmological horizon which appears in the case $\La > 0$, which is always much greater than the others for small $\La$).  In this case $f(r_h)=0$ and $f'(r_h)\neq 0$. For $a_0$ near the event horizon radius $r_h$, we have
\be
\sigma_0=-\frac{3}{4\pi r_h}\sqrt{f'(r_h)}\sqrt{a_0-r_h}+O(a_0-r_h)^{3/2},
\ee
\be
p_0=\frac{1}{8\pi}\frac{\sqrt{f'(r_h)}}{\sqrt{a_0-r_h}}+O(a_0-r_h)^{1/2},
\ee
so that as $a_0$ approaches $r_h$ the total amount of exotic matter $\Omega$ tends to vanish, but the pressure takes unlimitedly large values. Instead, when $|Q|=|Q_c|$, $f(r_h)=f'(r_h)= 0$ and  $f''(r_h)\neq 0$,  we have
\be
\sigma_0=-\frac{3}{4\sqrt{2}\pi r_h}\sqrt{f''(r_h)}(a_0-r_h)+O(a_0-r_h)^{2},
\ee
\be
p_0=\frac{\sqrt{2}}{8\pi}\sqrt{f''(r_h)}+O(a_0-r_h),
\ee
so that the amount of exotic matter $\Omega$ can be made as small as desired, keeping the pressure finite,  by taking $a_0$ near  $r_h$. One can also verify that if $|Q|>|Q_c|$ then the original metric includes no horizons, and the amount of exotic matter has a minimum for $a_0$ slightly greater than $r_{min}$, while the pressure for this $a_0$ remains finite.

In Figs. \ref{en} we have plotted the total amount of exotic matter for different choices of the  parameters (always with $\alpha > 0$ and $\La=0$), in the case $|Q|\leq |Q_c|$, so that at least one horizon exists. In each figure, when $\alpha$ changes  we keep fixed the relation between the wormhole radius and the horizon; the range of possible values of charge is reduced as $\alpha$ increases. We can see that: 1) for a given value of $\alpha$ and a fixed radius of the wormhole, the total amount of exotic matter decreases as the charge is made larger; 2) for a given charge, the amount of exotic matter is reduced by increasing the value of $\alpha$; 3) the exotic matter present is reduced by placing the wormhole throat nearer the horizon.

\section{Discussion}

 We have studied five dimensional spherically symmetric thin-shell wormholes in Einstein--Maxwell theory with the addition of a Gauss--Bonnet term. We have analyzed the  mechanical stability under perturbations preserving the symmetry, and evaluated the total amount of exotic matter and  related its behavior with that of the pressure of the shell, as a function of  the parameters of the model and of the theory. 

For null $\alpha$ our paper extends to five dimensions the analysis made in previous works \cite{poisson,eilobo} in four dimensional gravity. But when $\alpha\neq 0$ new interesting results are obtained.  We have found that the inclusion of the quadratic correction (Gauss--Bonnet term) allows for much more freedom in the choice of the configurations to render them stable. In particular, a central positive feature of the inclusion of the quadratic correction is that, differing from the general relativity case ($\alpha=0$), considerably larger regions of stability appear, even for vanishing charge; in this sense, we should emphasize that values of the parameter $\beta^2$ positive and  smaller than unity are now possible, while in the case  $\alpha=0$ this  could be achieved only with the aid of large values of  charge ($Q$ close to $\sqrt{3}M/\pi$). 

Regarding the amount of exotic matter $\Omega$ and pressure, the analysis shows  that, in the case that an event horizon exists in the original manifold from which we started our thin-shell construction,  $\Omega$ can be minimized by choosing a wormhole radius near the event horizon. In general this is correlated with an unlimited increase of the pressure, except if the parameters are chosen so that the inner and outer horizons  coincide: in this case the exotic energy can be reduced as desired, keeping the pressure finite. Besides, we have shown that as the Gauss--Bonnet term is made more relevant by increasing the parameter $\alpha$, the amount of exotic matter results to be substantially reduced.

\section*{Acknowledgments}

This work has been supported by Universidad de Buenos Aires (UBA) and CONICET. We wish to thank G. Giribet for valuable discussions and suggestions.

\end{document}